# AI Slop or AI-enhancement? Student Perceptions of AI-generated Media for an English for Academic Purposes Course

David James Woo [a], Deliang Wang [b] *, Kai Guo [c]

[a] Everwrite Limited, Hong Kong, China
[b] Faculty of Education, The University of Hong Kong, Hong Kong, China
[c] Faculty of Education, The Chinese University of Hong Kong, Hong Kong, China

* **Corresponding author**

- Postal address: Room 525, Meng Wah Complex, The University of Hong Kong, Pokfulam Road, Hong Kong, China
- Email address: wdeliang@connect.hku.hk
- Phone: +852 54879580

## STATEMENTS AND DECLARATIONS

### Availability of data and materials

The datasets used and/or analysed during the current study are available from the first author on reasonable request.

### Competing interests

The authors declare that they have no competing interests.

### Funding

This research received no specific grant from any funding agency in the public, commercial, or not-for-profit sectors.

### Authors' contributions

**DJW** conceptualized and designed the work; acquired, analyzed and interpreted data; and drafted the work and revised it. **DW** acquired, analyzed and interpreted the data; and drafted the work and revised it. **KG** drafted the work and reviewed it. All authors read and approved the final manuscript.

## AUTHOR BIOGRAPHY

***Dr. David James Woo*** is a secondary school teacher. His research interests are in artificial intelligence, natural language processing, digital literacy, and educational technology innovations. Email address: drdavidjameswoo@gmail.com; ORCID: https://orcid.org/0000-0003-4417-3686

***Dr. Deliang Wang*** is an incoming postdoctoral researcher in the Faculty of Information at the University of Toronto. His research primarily focuses on AI in education (AIEd), with a particular emphasis on understanding and optimizing AI-supported learning experiences within STEAM and language education contexts. His recent work has been published in prestigious international peer-reviewed journals, including *Educational Research Review, British Journal of Educational Technology, IEEE Transactions on Learning Technologies, Education and Information Technologies, Educational Technology & Society, International Journal of Educational Research, Journal of Science Education and Technology,* and *the International Journal of Artificial Intelligence in Education*. As of April 2026, his research has been cited over 1,300 times according to Google Scholar. Email address: wdeliang@connect.hku.hk; ORCID: https://orcid.org/0009-0008-6488-0234

***Dr. Kai Guo*** is an RGC Junior Research Fellow and a Research Assistant Professor in the Department of Curriculum and Instruction at The Chinese University of Hong Kong. His research interests include technology-enhanced language learning and second language writing. His recent publications have appeared in international peer-reviewed journals such as *Computers & Education*, *The Internet and Higher Education*, *British Journal of Educational Technology*, *Computer Assisted Language Learning*, *Interactive Learning Environments*, *Language Learning & Technology*, *System*, *TESOL Quarterly*, *Assessing Writing*, and *Innovation in Language Learning and Teaching*. Email address: kaiguo@cuhk.edu.hk; ORCID ID: http://orcid.org/0000-0001-9699-7527

# AI Slop or AI-enhancement? Student Perceptions of AI-generated Media for an English for Academic Purposes Course

## Abstract

Artificial intelligence (AI) retrieval-augmented generation (RAG) tools now enable educators to transform course materials into diverse multimedia at scale. However, it remains unclear whether such AI-generated content functions as a pedagogical scaffold or *AI slop*: high-volume, low-quality material. This innovative practice paper reports on the development, implementation, and evaluation of teacher-prompted, AI-generated supplemental materials in an English for Academic Purposes (EAP) course at a Hong Kong community college. Using primarily Google NotebookLM, the instructor generated videos, podcasts, infographics, and individualized feedback reports from course materials and student work for 106 English as a Foreign Language learners. An explanatory sequential mixed-methods design comprising a survey, semi-structured interviews, and correlation analysis with academic scores was employed to examine students' preferences, perceptions, and learning outcomes. Findings are framed through the Technology Acceptance Model and Cognitive Load Theory. Students rated the materials highly for perceived usefulness and ease of use, and preferred assessment-linked content presented in visual and multimodal formats, particularly videos and infographics. Video preference correlated positively with academic performance. However, higher cognitive load was negatively associated with course grades, indicating that material complexity must be carefully calibrated. Notably, some lower-performing students independently adopted the materials as remedial scaffolds. The practice demonstrates that RAG tools enable scalable, personalized feedback that would be less feasible through traditional methods. When aligned with student goals and cognitive principles, teacher-prompted AI generation can meaningfully enhance the EAP learning ecosystem rather than producing AI slop.

**Keywords:** AI-generated materials; English for Academic Purposes; multimedia learning; cognitive load; technology acceptance; retrieval-augmented generation

## Introduction

Artificial intelligence (AI) capabilities have advanced beyond text generation to include automated generation of multimedia content (e.g. audio; images; and video; Wang et al., 2025) and transformation of content through retrieval-augmented generation (RAG), which combines information retrieval with generative models to enhance content accuracy and relevance (Li et al., 2025). Tools such as Google's NotebookLM combine AI multimedia and RAG capabilities and allow educators to transform content into diverse multimedia at little or no cost (Wang, 2024). The integration of these tools into school and university pedagogical practices could constitute a curricular innovation if the proliferation of multimedia enhances student learning. On the other hand, it may merely lead to *AI slop*, high-volume, low-quality, AI-generated materials that clutter learning management systems (LMSs) (Jones et al., 2025).

Distinguishing *enhancement* from *slop* requires attending to at least three theoretical dimensions. First, the Technology Acceptance Model (TAM) predicts that students will engage with supplemental materials only to the extent that they perceive them as useful and easy to use (Granić & Marangunić, 2019). In high-stakes academic pathways, these perceptions may be shaped by whether materials align with assessed tasks. Second, Cognitive Load Theory (CLT) warns that the volume and complexity AI tools can produce risk imposing unnecessary cognitive demand, exhausting the processing capacity EFL learners need for deeper understanding, even as the germane load that builds schemas and integrates knowledge should be preserved (Sweller, 2011). However, whether such engagement translates into higher grades may depend on how performance is assessed and scored. Third, Mayer's (1997) Cognitive Theory of Multimedia Learning suggests that formats like video and infographics may lower comprehension barriers by distributing processing across visual and auditory channels, but the coherence principle cautions that extraneous elements can distract rather than support learning. These dimensions interact: a cognitively manageable, visually engaging resource may still go unused if students see it as irrelevant to their grades. On the other hand, a demanding text report may be valued if it delivers individualized, grade-relevant feedback.

Student perceptions and performance can signal whether AI-generated multimedia functions as a pedagogical scaffold or AI slop. In the context of English for Academic Purposes (EAP) instruction for English as a Foreign Language (EFL) learners, empirical evidence is needed not only to evaluate effectiveness, but to theorize what AI integration enhances (Kohnke, 2024). Thus, this innovative practice paper reports Hong Kong community college EFL students' preferences and perceptions of AI-generated multimedia materials developed and implemented by their EAP course instructor (the first author) to supplement instruction. Reporting preferences and perceptions along with learning outcomes, and drawing on TAM, CLT and multimedia learning principles**,** we argue that the AI-generated multimedia enhanced the course ecosystem in terms of perceived relevance and cognitive accessibility. Importantly, the innovative practice demonstrates how free, RAG-based AI tools enable scalable personalization of feedback that would be less feasible with traditional methods.

## Materials and Methods

### Context

The practice was implemented during the fall 2025-26 semester in a Hong Kong Community College that delivers two-year associate degree programs. Associate degree students may receive offers for undergraduate programs in Hong Kong universities based on their academic performance in the College. Students at the College include those who were not offered undergraduate program places based on their university entrance exam results from secondary school; and students who were offered undergraduate places, but sought to improve their academic performance and receive more preferred undergraduate admission offers. In that way, the range of learner diversity could be wide.

The College delivers a compulsory EAP course for its associate degree students, comprising 36 contact hours and 24, 90-minute lessons. The course aims to increase students' flexibility and precision in using written and spoken English for academic settings, to build their interpretive skills when analyzing tabular and graphic data and to enhance their critical selection of research for academic assignments. Its intended learning outcomes (ILOs) are for students: to use the language of argumentation and discussion (ILO1); to describe charts, graphs, and tables effectively (ILO2); to utilize the language of comparison and contrast (ILO3); to write a simple research report using academic conventions (ILO4); and to blend secondary research into academic discussion using the APA style of referencing (ILO5).

During the course (see Supplemental Material for course outline), students would be given two homework tasks: writing an adapted IELTS academic writing task 1; and writing a discussion paragraph for a research report. Summative assessments include Assignment 1 (CA1): a task adapted from IELTS academic writing task 1 (30% of total score); Assignment 2 (CA2): a 3,000-4,000 word group research report (30% of total score) developed from a questionnaire; and Assignment 3 (CA3): a group oral presentation of that report (30% of total score). Course instructors should also assess students on class participation (CA4) (10% of total score), of which punctual attendance would be a key consideration. Additionally, instructors should normalize students' overall scores for each class and students' scores could be further moderated by College administrators.

Compulsory course materials include a set of worksheets with answer keys for each lesson. Instructors could develop supplemental course materials. In that way, the first author used AI tools to generate supplemental material based on the compulsory course material and other sources.

**Participants**

The AI-generated supplemental materials were received by 106 second year associate degree students comprising three classes taught by the first author.

**Description of Supplemental Materials**

Table 1 lists the supplemental materials by media type. AI-generated transcripts, necessary for AI tools to convert video and audio recordings into other media, were not intended to be supplemental materials and are not listed. Only materials generated within the scope of study, from September 2025 to December 5, 2025, before the end of course, are included in the table.

The material listed in Table 1 is further organized by curriculum area, which refers to what the media was intended to supplement: for instance, lesson worksheets refer to the verbatim content presented on the compulsory worksheets. Media intended to supplement derivative activity based on lesson worksheets, such as in-class activities or homework tasks, are listed separately. ILOs are also aligned with each material.

Table 1 provides a description and input sources for each media type for each curriculum area. Moreover, the table provides the total number of media for each curriculum area that was produced during the study period and an allocation per unit if applicable. Finally, the table lists the AI tool(s) used to generate the material and whether any custom prompts (e.g. to remove student names; to specify criteria for AI generation) were used to generate the material. Otherwise, the AI tool could generate material based on the given input sources and its default settings.

The supplemental materials were generated primarily with Google NotebookLM (https://notebooklm.google.com/), an AI-powered, RAG tool: as seen in Figure 1, on the left hand side a user adds sources such as Google Docs, PDF files, websites and YouTube videos; the user can then interact with those sources through a chatbot interface in the center console; and on the right hand side the user can transform those sources into customizable audio and video (e.g. either around six minutes or two minutes in length), written reports, flashcards, mind maps and quizzes.

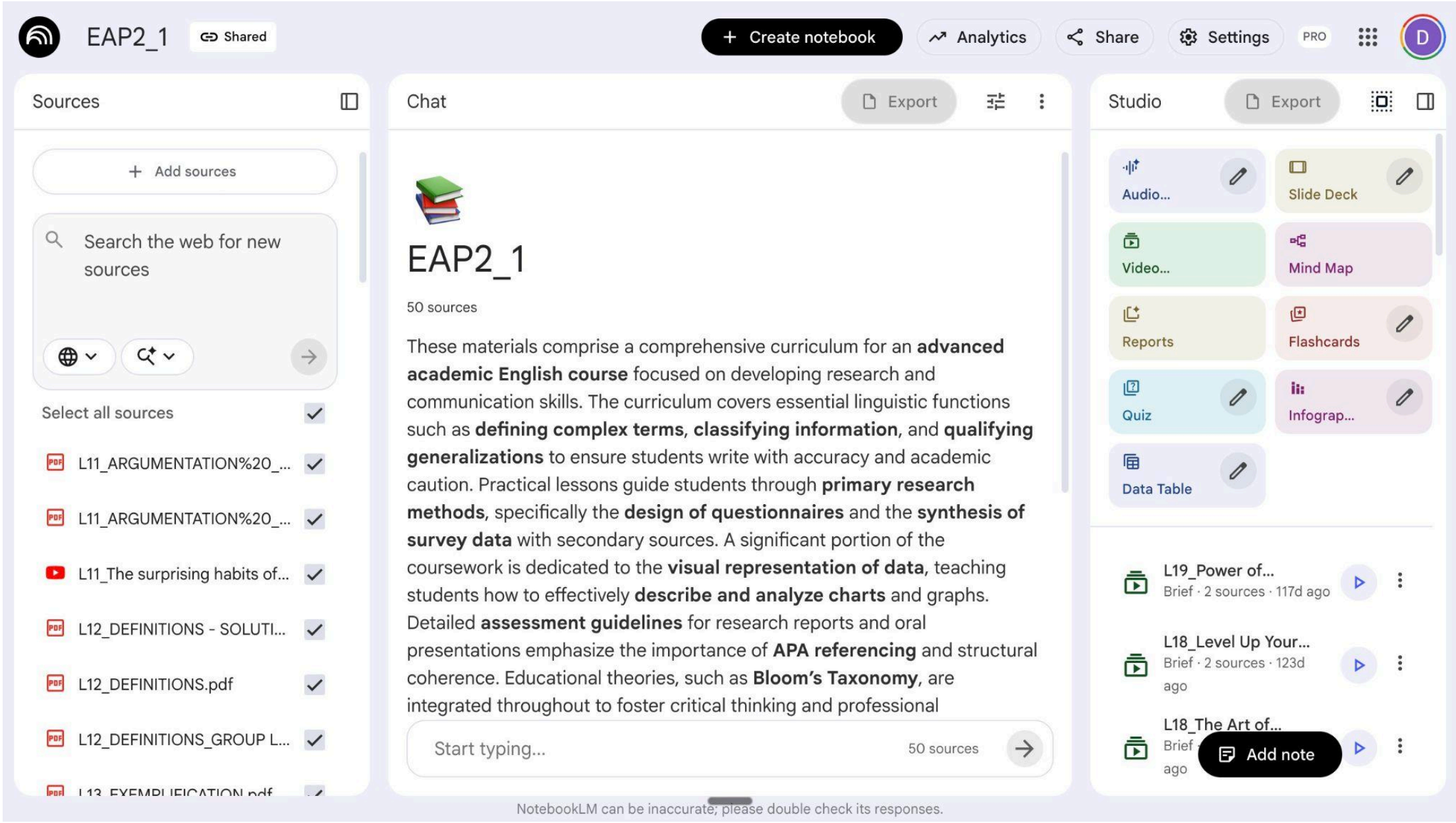

**Figure 1.** Google NotebookLM interface circa March 2026

*Figure 1 alt text: a March 2026 screenshot of the Google NotebookLM interface, showing a sources console on the left-hand side, a chatbot in the center console and a studio console on the right-hand side*

The first author used the free version opportunistically: first, Google updated its tool during the semester, releasing new features, such as importing sources directly from Google Drive (October 2025) and allowing prompts to customize video styles (November 2025). Features such as generating infographics, slide decks and datasets were released after the study period. Second, Google constrained production of media to free version users, for instance, limiting a user to generate three videos maximum in a 24-hour period.

**Table 1.** An overview of supplemental materials by media type

| Media Type | Curriculum Areas | Intended learning outcomes (ILOs) | Description | Input Sources | Total Quantity (Allocation per unit) | AI Tools | Custom Prompt |
|---|---|---|---|---|---|---|---|
| Infographic | Assignment 1 | 2, 3 | Three points on how to improve scores | Assignment 1 reports | 1 | Grok | Yes |
| Infographic | Lesson worksheets | 1, 2, 3, 4, 5 | An overview of core ideas from a lesson's worksheets | Lesson worksheets with answer keys | 14 | Gemini<br>Grok<br>Napkin.ai | Yes |
| Podcast | Lesson worksheets | 1, 2, 3, 4, 5 | An overview of core ideas from a lesson's worksheets | Lesson worksheets with answer keys | 2 (1 per lesson) | Google NotebookLM | No |
| Report | Assignment 1 | 2, 3 | Class overall performance summary with grade distribution, detailed analysis by assessment criterion, exemplary performance and key strengths, and summary of findings | Students' individual Assignment 1 score sheets with comments | 4 (1 per class) | Google NotebookLM | No |
| Report | Homework 1 | 2, 3 | Class level feedback, and individual student feedback reports for specified assessment criteria | Students individually written homework assignments | 3 (1 per class) | Google NotebookLM | Yes |
| Report | Homework 2 | 1, 3, 4, 5 | Class level feedback, and individual student feedback reports for specified assessment criteria | Students individually written homework assignments | 2 (1 per class) | Google NotebookLM | Yes |
| Video | Assignment 1 | 2, 3 | Key elements of a high scoring Assignment 1 | Students' individual Assignment 1 score sheets with comments | 1 | Google NotebookLM | Yes |
| Video | Assignment 2 | 4 | Overview of instructor's feedback for revising research report questionnaires | Instructor's screen-recorded feedback while completing each group's research report questionnaire | 3 (1 per class) | Google NotebookLM | No |

| Video | Homework 1 | 2, 3 | Class level feedback for specified assessment criteria, citing examples from specific students | Students individually written homework assignments | 1 | Google NotebookLM | Yes |
|---|---|---|---|---|---|---|---|
| Video | Lesson worksheets | 1, 2, 3, 4, 5 | An overview of core ideas from a lesson's worksheets | Lesson worksheets with answer keys | 24 (1 per lesson) | Google NotebookLM | No |

## Implementation

The supplemental materials were placed in a shared Google Drive folder called Supplemental Materials. They were further organized into subfolders according to curriculum area (e.g. Assignment 1; Homework 1; Homework 2) and then by class. Materials meant for individual students or groups had specific names in the file name. A link to the Supplemental Materials folder was in the College's LMS, which also stored the core compulsory materials.

The instructor presented relevant materials during lessons, for instance, showing a lesson video or a homework report or video generated in Google NotebookLM (see Figure 2) when providing homework feedback or an Assignment 1 report when providing Assignment 1 feedback; and sent links to specific materials through LMS announcements and emails.

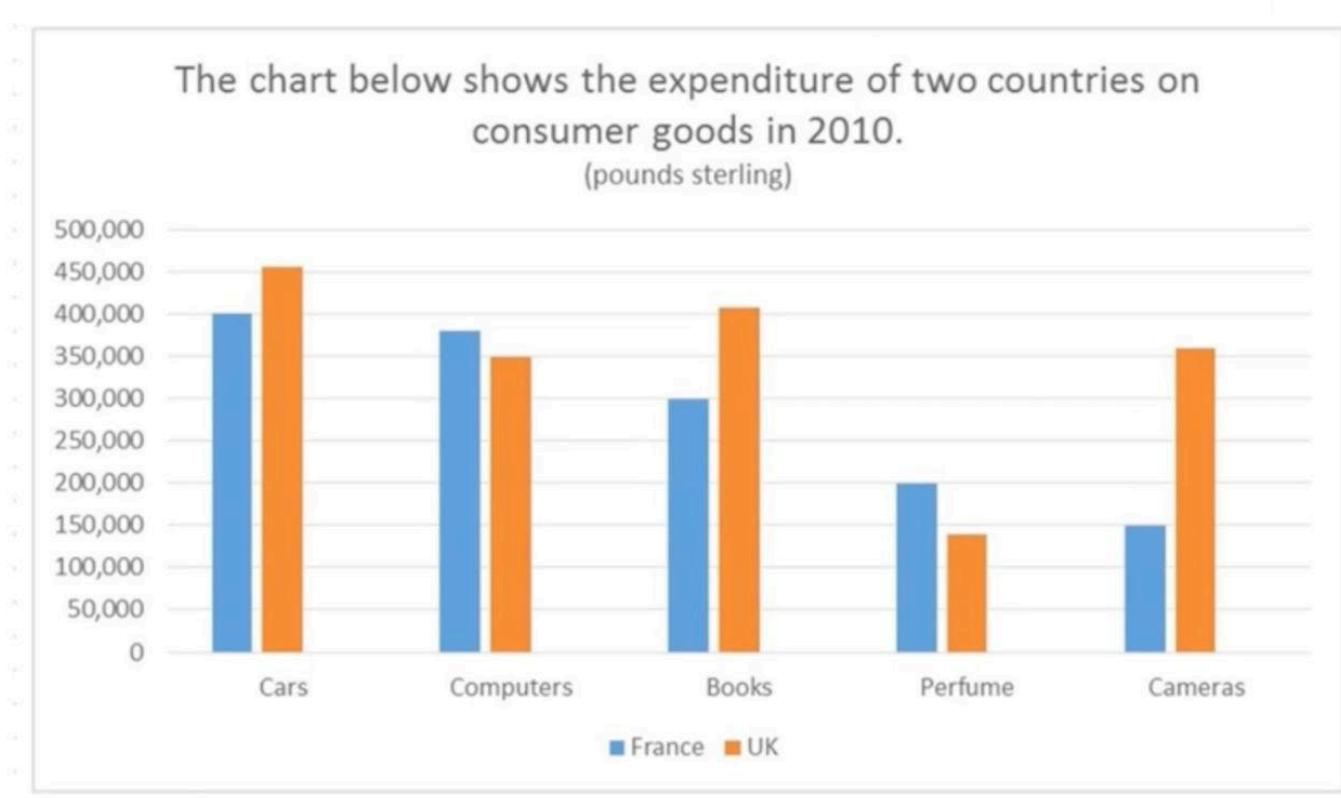

**Figure 2.** A screenshot from a video presenting class-level feedback from Homework 1

To collect students' preferences and perceptions of the AI-generated supplemental materials, we employed an explanatory sequential mixed-methods design (Creswell & Plano Clark, 2018) where a qualitative phase follows a quantitative phase with the purpose to elaborate specific quantitative results.

## Survey

For the quantitative phase, we prepared a 35-item Google Form survey. We theoretically approached students' preferences and perceptions through the TAM (Wang et al., 2024) comprising the dimensions of perceived usefulness (PU) (i.e. evaluates how helpful the materials are for learning) and perceived ease of use (PEU) (i.e. evaluates the ease of accessing and operating the materials). To measure if the AI materials made concepts easier or harder to process mentally, we also sought to measure students' cognitive load (CL) (i.e. evaluates the mental effort and frustration involved in understanding the materials) (Hwang et al., 2013) and germane load (GL) (i.e. evaluates the extent to which materials promote deep thinking and knowledge integration) (Sweller, 2011). Thus, we designed 15, 5-point Likert-Scale (1 - strongly disagree; and 5 - strongly agree) items querying students' frequency of use, PU and PEU of supplemental materials; 12 Likert-Scale items related to CL and GL; six demographic items (e.g. gender; program year; public housing as an indicator of socio-economic status) including whether a student had independently accessed supplemental material; and two follow-up items, including whether students would be willing to be interviewed.

In preparing the survey items, we purposefully omitted AI from the questions and referred to materials as supplemental. In that way, we wished students to focus on the material contents, not about the AI-generated nature of them. Furthermore, the survey was piloted to ensure students could understand questions and complete the survey within five minutes.

The survey was sent to 106 students via the LMS and email. The instructor announced the survey to students in each class during the penultimate lesson of the course and gave students time to complete the survey during the lesson. The instructor also informed the students of the purpose and benefits of the study, the anonymization of their data, students' right to decline participation at any time, and their rights as participants.

We quantified the subjective constructs, specifically calculating the mean scores of four core dimensions from the Likert-scale survey results. Additionally, we converted students' rankings into continuous preference scores. Rank 1 (used most) was scored as 4 points, and Rank 4 (used least) was scored as 1 point. A higher score indicates a stronger preference for, or reliance on, that specific material/media.We report those and other descriptive statistics from students' survey results.

To explore students' preferences for AI-generated supplemental materials, their subjective perceptions, and the relationship with learning outcomes, we operationalized student learning outcomes in terms of their academic performance for the course assignments and overall course score. In that way, the instructor had collected students' three summative assessment and class participation scores (CA1 to CA4) which comprise the overall score (Total Score). Students' overall scores showed a normal distribution. Next, we would calculate the correlation between perceptions, preferences and scores by matching the survey data with students' scores, and utilizing Pearson Correlation Coefficients to test the linear relationships among variables.

**Interview Protocol**

For the qualitative phase, we invited students for a structured, 5-minute interview between the individual student and the instructor. The interview protocol focused on asking students five questions based on the supplemental material they used most frequently. The questions were developed with experts in Multimedia Learning Theory (Mayer, 1997) and Cognitive Load Theory (Sweller, 2011). We recorded and transcribed the interviews.

To explain students' preferences and perceptions, we analyzed interview transcripts using directed content analysis (Hsieh & Shannon, 2005). First, we coded students' responses according to media types and curriculum areas so that we could easily corroborate survey trends with student accounts. Following that, we applied an inductive thematic analysis (Braun & Clarke, 2006) to identify patterns of meaning within and across the media types and curriculum areas. This involved reviewing the coded excerpts to generate broader themes to capture underlying reasons. We report verbatim quotes from the interviews.

To triangulate student survey and interview data, the first author contributed informal observations, recorded retrospectively, reflecting the instructor's in-class implementation experience and student responses. These practitioner insights merely corroborate or extend the other findings.

## Results and Discussion

Out of 106 students, 45% (n=48) responded to the survey. 10 students indicated in the survey that they did not access or view any of the supplemental material so their surveys were not further analyzed. We integrated and quantitatively analyzed the remaining students' survey data (N=38) and academic score data (N=36). 24 students identified as female, 13 as male and 1 non-binary.

15 students had indicated a willingness to be interviewed and three students were successfully interviewed. Each student had indicated a different type of most frequently used media based on curriculum area, either Assignment 1 infographics, reports and videos; lesson worksheet videos; or homework reports and videos.

The results and discussion are organized as practical benefits and challenges of teacher-prompted, AI-generated supplemental materials in an EAP course. Benefits are reported first, followed by challenges, with survey findings, correlation data, interview accounts and instructor observations integrated within each.

### Practical Benefits

Students perceived the supplemental material positively overall. As shown in Figure 3, they gave high scores for PU (Mean = 3.91/5) and PEU (Mean = 3.75/5). These scores indicate that students valued these AI-generated supplemental materials and found them easy to use. Additionally, the relatively high GL score (Mean = 3.53/5) indicates that the supplemental materials could successfully stimulate students to engage in deep reflection and connect new knowledge with their existing knowledge framework.

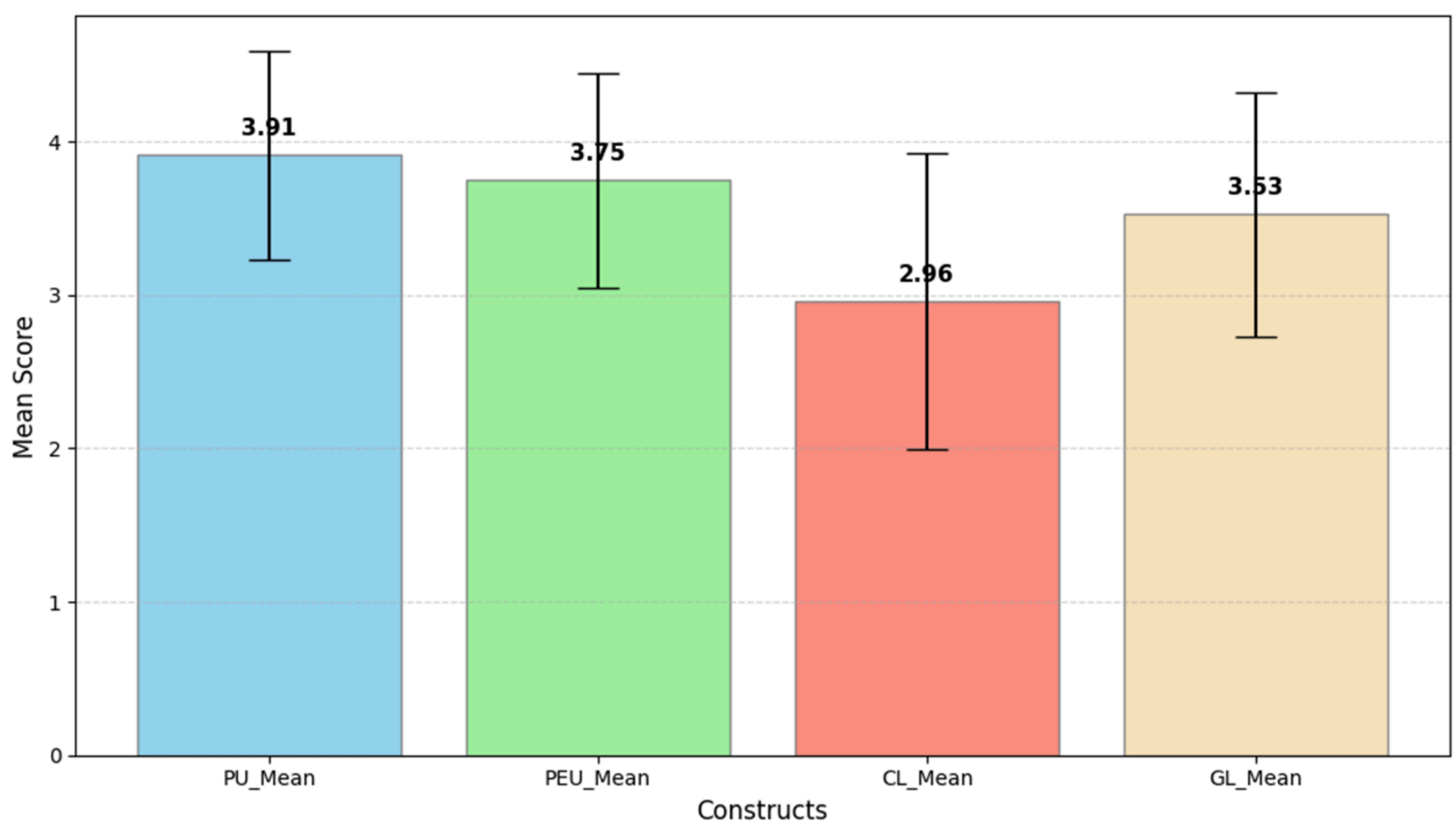

**Figure 3.** Descriptive statistics of students' perceived usefulness, perceived ease of use, cognitive load, and germane load.

Students most frequently used materials related to assessments. As presented in Figure 4, "Assignment 1 infographics, reports and videos" was dominant, ranked first by nearly half (n=18) of students, followed by "Homework reports and videos." Conversely, "Lesson worksheet videos," which are most indirectly related to assessments, had the lowest usage rate. Interviews also suggest students preferred materials that directly contributed to grades. For instance, one student who most frequently used the Assignment 1 Performance Report said its "detailed information has enabled me to better understand where more attention and improvement are needed." The student who most frequently used the homework reports said, "I think the information is long enough for me to improve." Even the student who most frequently used lesson worksheet videos cited the video on designing survey questionnaires for Assignment 2 as particularly memorable.

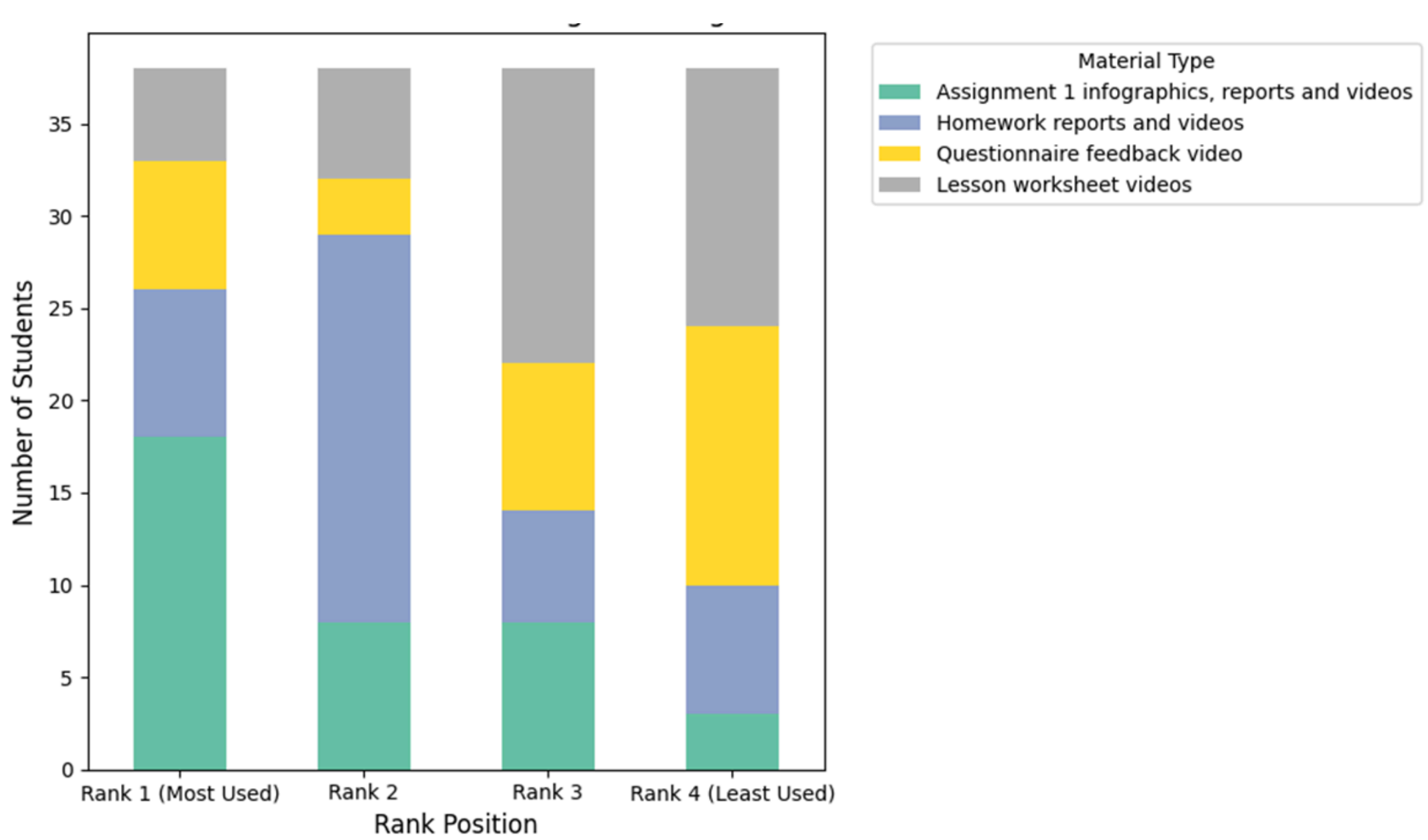

**Figure 4.** Distribution of material usage rankings.

From the instructor's perspective, the prioritization of assessment-linked materials emerged during the course, influenced by Google Notebook LM's daily generation limits for videos and infographics. Moreover, prior to the course, colleagues had indicated that students in the course would be grade-oriented; and that was corroborated in lessons when students appeared to pay greater attention when the instructor introduced grade distribution tables and individual feedback reports, compared to other material types (e.g. lesson worksheet videos).

Visual and text-combination and text formats were the most popular media types. As shown in Figure 5, "Infographic" and "Report" were chosen as the most frequently used media by the majority of students. In contrast, the audio-only "Podcast audio" was extremely unpopular, with 18 students ranking it last.

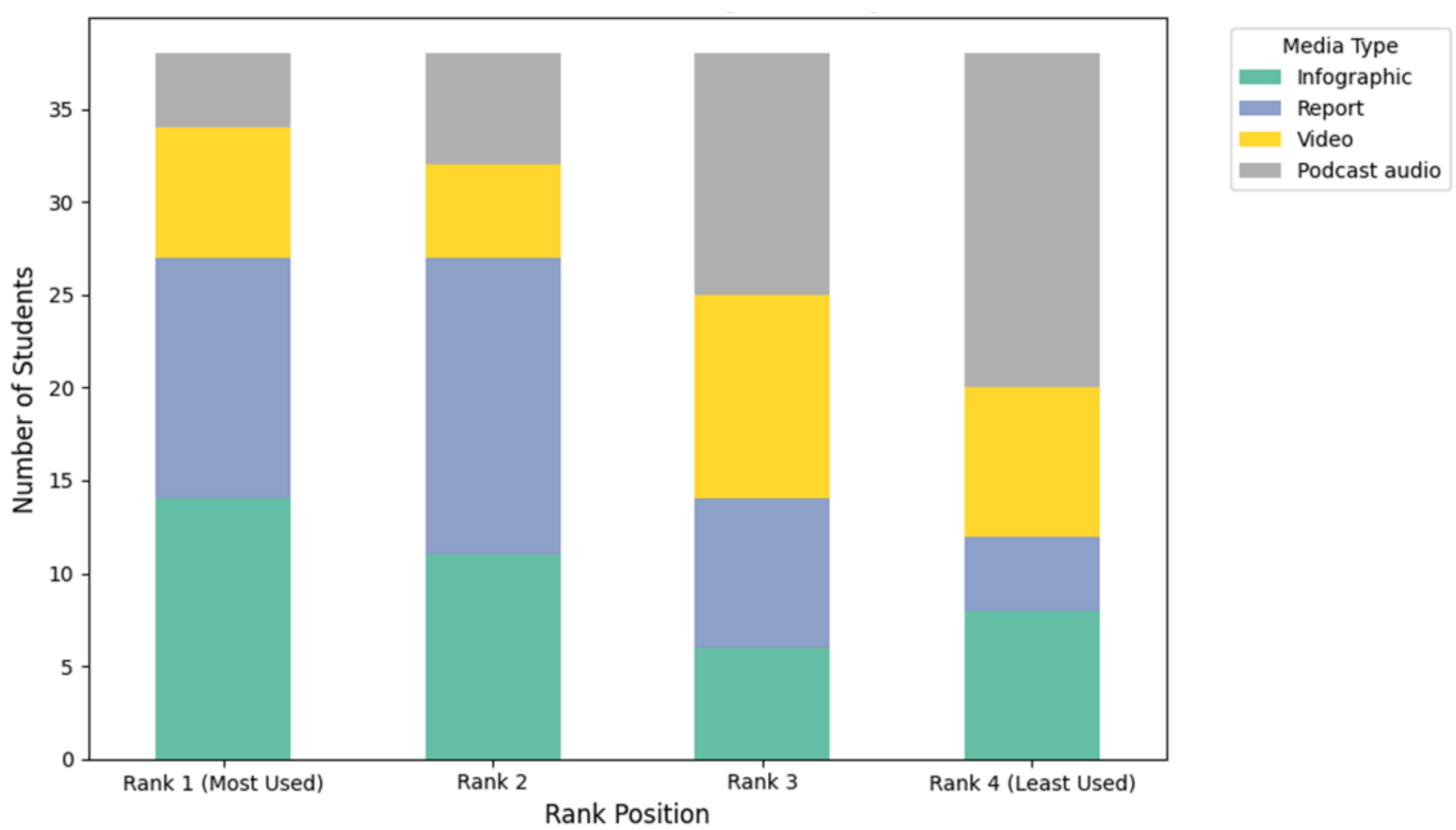

**Figure 5.** Distribution of media usage rankings.

All three interviewees identified visual or structural features as important for comprehension. The student who most frequently used Assignment 1 Performance Report cited a table as a most beneficial feature because it, "clearly lists the quantities of different grades," and easily helped her, "find the information I want to see, such as knowing the average score." Likewise, the student who most frequently used homework reports said the tables in the reports were fine. All three students reported their materials (i.e. reports and videos) were of sufficient length and had no suggestions to improve their materials' structure.

Importantly, as shown in Figure 6, students who preferred using "Video" showed comprehensive and highly positive feedback: they found the materials extremely useful (PU correlation: 0.460) and easy to use (PEU correlation: 0.399); their deep thinking was significantly stimulated (GL correlation: 0.499); and most importantly, video preference correlated positively with total academic scores (0.283), especially CA3 (0.321). One student whose most-used material was a lesson worksheet video stated in videos, "the visual elements are more prominent..and captures our attention more."

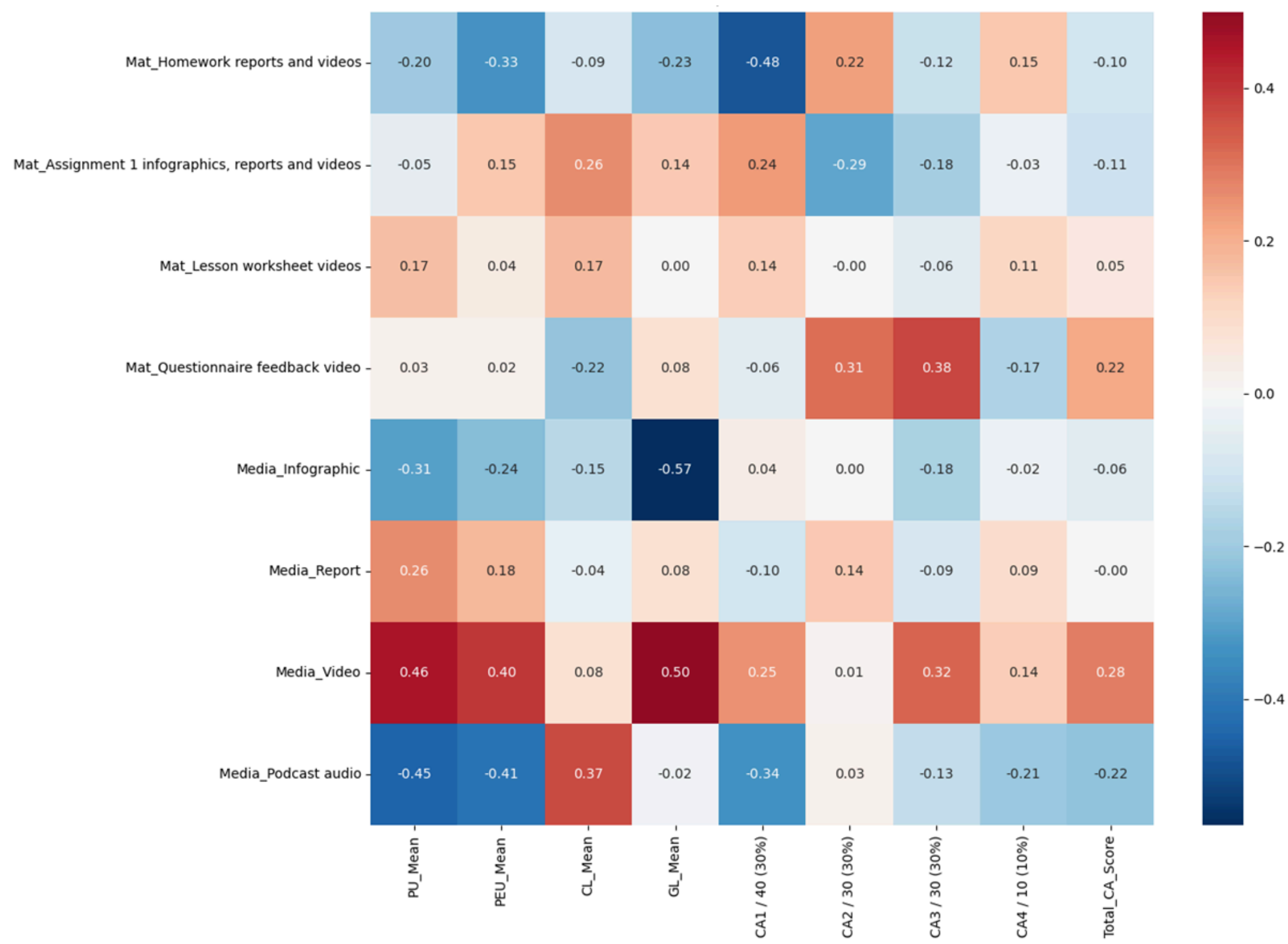

**Figure 6.** The correlation between material/media preferences and learning outcomes.

However, the instructor observed that student attentiveness to lesson worksheet videos appeared to decline over the course of the semester, which he attributes to the videos being not directly tied to academic performance; and their novelty diminishing. In response, the instructor reduced the lesson worksheet video length from approximately six minutes to two minutes, a decision reinforced by a colleague's observation that the original length appeared excessive. This adjustment suggests AI-generated media may require ongoing revision to sustain engagement.

In sum, the student preferences indicate that AI-generated materials were perceived as enhancements at the intersection of goal alignment and multimodal design. Materials that were associated with assessments and presented in visually structured formats received the highest usage and preference ratings. This suggests that for EFL learners in high-stakes academic pathways, AI-generated content has value contingent upon perceived alignment with goals and multimodality.

**Practical Challenges**

The CL (Mean = 2.96/5) was at a moderate level, showing that the materials presented some challenges. Significantly, CL was negatively correlated with scores across all stages (CA1–CA4) and the total score (correlation with Total_CA = −0.316), as shown in Figure 7. This pattern is consistent with the prediction from Cognitive Load Theory that students who found the materials more effortful and difficult tended to achieve lower final grades, though the direction of this relationship cannot be determined from the present data — it is equally plausible that lower-proficiency students experienced higher cognitive load.

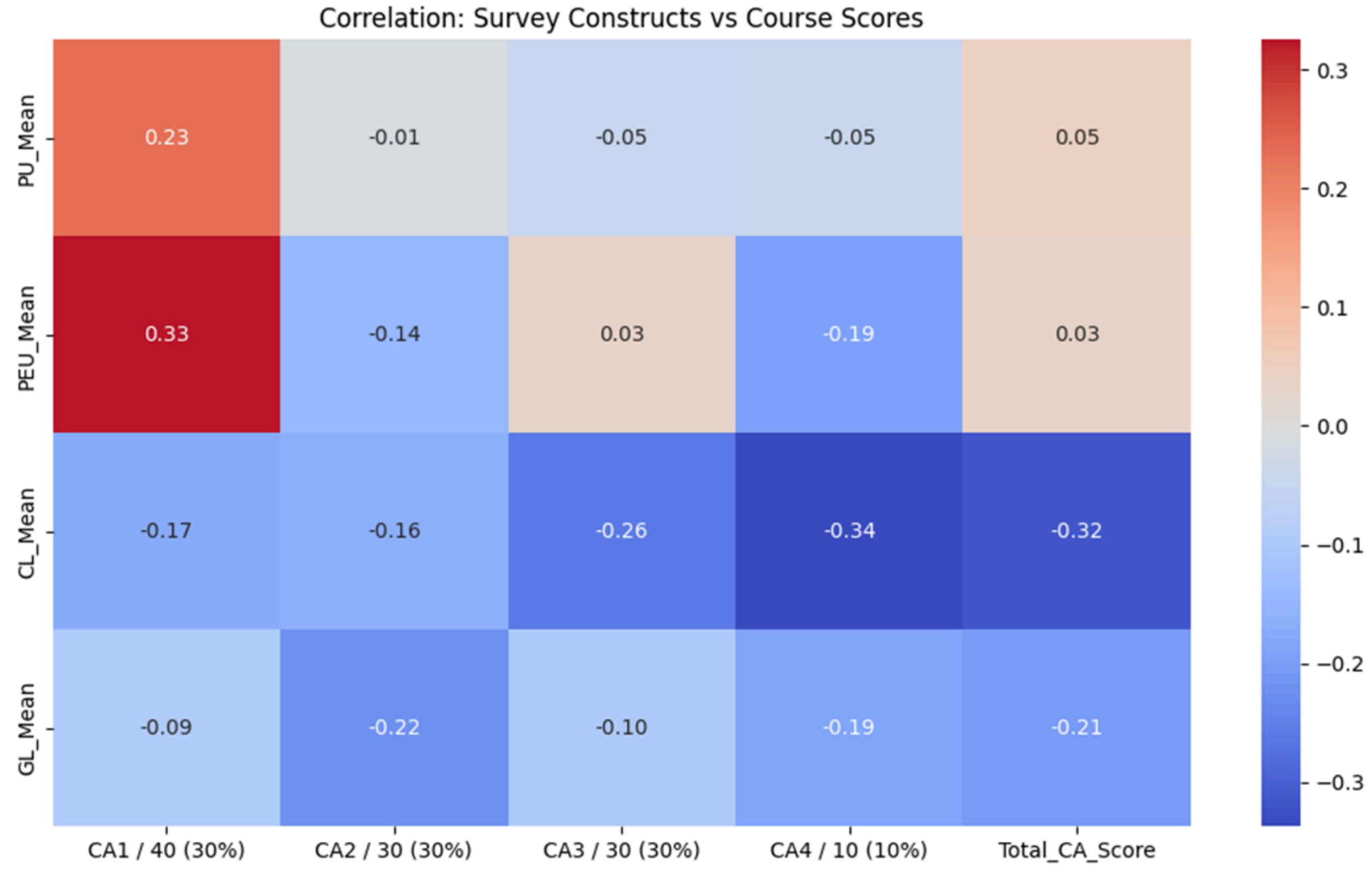

**Figure 7.** The correlation between survey constructs and course scores.

Another challenge is that some students struggled in the course and exhibited remedial learning behavior. These students relied heavily on "Homework reports and videos" and showed a strong negative correlation with CA1 scores (-0.478) (see Figure 6). The implication is that these students who struggled in CA1 turned to homework feedback materials to catch up, rather than the materials causing low scores. One student supported this pattern, engaging primarily with their own individual feedback rather than class-level summaries and confirmed the report contained sufficient information to support improvement.

Notably, students who adopted a remedial pathway did so independently, without instructor guidance as, at the time, the instructor was unaware that this pattern was emerging. In hindsight, the instructor reflects that explicitly orienting all students to the materials and their value at the start of the course may have broadened and deepened self-directed remedial behavior.

A related challenge is the College's assessment policies that may not reward students' efforts. Surprisingly, GL (promoting deep thinking) showed a certain degree of negative correlation with the total score (-0.206) and CA2 (-0.223) (see Figure 7). This may imply that under the current assessment mechanism, students who engage excessively in deep reflection and knowledge linking (high GL) might not necessarily have an advantage in specific standardized tests (such as CA2). We can understand this because students' academic performance is normalized and moderated for the course. If students' scores were exclusively criterion-referenced, allowing more students to score higher grades, we may observe a different result from the correlation test.

**Limitations**

The sample size and survey response rate were relatively low. Institutions that attempt to scale the use of teacher-prompted, AI-generated supplemental materials for EAP, to course level, for instance, could consider increasing sample size and response rate by adding items to existing courses and teaching evaluation surveys typically delivered at the end of a course. The interviews provide insight into individual differences for preferences and perceptions, but can not be generalized to the survey sample.

Our understanding of any significant relationship between students' perceptions or preferences and their EAP performance is limited to norm-referenced academic performance. In that way, further research should be carried out in higher education institutions which feature criterion-referenced academic performance for EAP courses so that, potentially, all students in a class can score As.

The study was also cross-sectional and students were surveyed before the official end of the course and before teacher-prompted, AI-generated supplemental materials could be generated to provide individualized feedback for CA2 and CA3. In addition, the development of AI-generated supplemental materials was not uniformly distributed for curriculum materials in terms of number of media types and number of items. In that way, further study might sample students' preferences and perceptions during a course and after a course with a more equal distribution of materials and media types. Importantly, we did not collect any process-oriented data, specifically of how students reviewed the supplemental materials and integrated them into their learning process. In that way, future research may want to collect observational data such as screen recordings, investigating the processes by which students integrate these materials into their learning. Future research could also directly compare AI-generated materials with traditional teacher-created equivalents.

**Lessons Learned**

The high PU, PEU and GL ratings indicate that teacher-prompted, AI-generated supplemental materials can function as effective learning scaffolds in this study's community college context. Students perceived the usefulness of the materials because material use was not only perceived as personally addressed to them but also tied to their goal of undergraduate admission based on academic performance. In that way, educators in similar contexts may generate materials that provide individualized feedback on formative and summative assessments that advance students towards their material goals. Individualized, AI-generated material may also address the learner diversity that we observed in our cohort, as some students struggled in the course and used AI-generated material to catch up.

The data strongly supports using "Video" and "Infographic" as the core media for supplementary materials, and video preference was most strongly associated with academic performance. This is not least because video can simultaneously engage visual and auditory senses, greatly lowering the barrier to understanding. If possible, educators in similar contexts may consider generating infographics or videos to provide assessment support and individualized feedback. Since interviewed students appeared satisfied with existing visual or structural features, educators who are novice AI users may not need to engineer custom prompts that could pose difficulties. However, the generation limitations of the AI tool (e.g. number of infographics or videos allowed to produce daily) would certainly be a constraint.

The moderate cognitive load mean score showed that the materials presented some challenges but did not make students feel extremely frustrated. Educators in similar contexts should be cautious about not overwhelming students by the breadth of information presented in AI-generated material. To avoid high cognitive load, educators could reduce the length of videos, the text on infographics and the length of reports, and calibrate the language in the materials to cohere with students' comprehension abilities. Additionally, educators should be mindful of organizing AI-generated materials in a coherent and efficient way for students to independently access them. They should not merely dump AI-generated material in a folder or send students frequent messages with AI-generated material.

Finally, since Cognitive Load (CL) has a clear negative correlation with final grades, instructors should identify students who find the “materials difficult to understand” early in the semester (e.g. after CA1) and provide direct guidance, as they are a high-risk group for ultimately failing to meet academic standards.

## Conclusion

Per this paper's title, students overwhelmingly perceived teacher-prompted, AI-generated media as AI-enhancement rather than AI slop. This enhancement operated primarily at the level of perceived value (i.e. materials were seen as personally useful and goal-aligned) and cognitive accessibility (i.e. multimodal formats like video and infographics reduced barriers to understanding); and we found students' video preference correlated positively with their academic performance for the course. Compared to traditional methods, the mechanism for the enhancement was AI's capacity for scalable, RAG-based personalization that allowed an instructor to provide differentiated feedback and multimodal resources to over 100 students. Furthermore, the relationship with academic performance was differential, suggesting that AI tools may function as performance accelerants for some learners and remedial scaffolds for others. Our innovative practice suggests that when thoughtfully aligned with student goals and cognitive principles, teacher-prompted AI generation can meaningfully enhance the EAP learning ecosystem.

## Supplemental Material

### A. Course Outline

| Lesson | Content |
|---|---|
| 1 | Introduction to the course<br>Research Report format |
| 2 | Selecting a topic for investigation |
| 3 | Identifying discourse features |
| 4 | Designing questionnaires |
| 5 | Describing charts, graphs, and tables 1 |
| 6 | Describing charts, graphs, and tables 2 |
| 7 | Describing charts, graphs, and tables 3 |
| 8 | Synthesizing ideas |
| 9 | Argumentation and discussion 1 |
| 10 | Assessment 1 (in-class writing test) |
| 11 | Argumentation and discussion 2 |
| 12 | Definitions |
| 13 | Exemplification |
| 14 | Qualifying generalizations |
| 15 | Comparison and contrast |
| 16 | Reduced clauses |
| 17 | Conciseness and clarity |
| 18 | Word families and parallel structures<br>Classification |
| 19 | Proofreading |
| 20 | Effective presentations |
| 21-24 | Assignment 3: Group Oral Presentation |

## B. Survey Items

| No. | Variable / Construct | Question / Item |
|---|---|---|
| 1 | Demographic | What is your gender? |
| 2 | Demographic | What is your programme year? |
| 3 | Demographic | Do you live in a public housing estate? |
| 4 | Demographic | What was your Hong Kong Diploma of Secondary Education (HKDSE) English grade level? |
| 5 | General Access (Filter) | Did you use any of (the instructor's) IELTS writing task 1 visual diagram generators on POE? |
| 6 | General Access (Filter) | Did you use any of (the instructor's) prompts for generating research questions or questionnaire items? |
| 7 | General Access (Filter) | Did you access or view any of the supplemental material provided in this course? |
| 8 | Usage Ranking (Behavior) | Please rank the following four types of materials based on how frequently you used them. (Rank 1 = Used Most, Rank 2 = Used Second Most, Rank 3 = Used Least) |
| 9 | Usage Ranking (Behavior) | Please rank the following four types of media based on how frequently you used them. |
| 10 | Perceived Usefulness (PU1) | The supplemental materials enriched my learning activity in this course. |
| 11 | Perceived Usefulness (PU2) | The supplemental materials were helpful to me in acquiring new knowledge. |
| 12 | Perceived Usefulness (PU3) | The format provided by the supplemental materials smoothed my learning process. |
| 13 | Coherence Principle | The length of the materials (e.g. podcast audio, report, video) negatively affected my ability to focus and learn. |
| 14 | Signaling Principle | The materials effectively highlight key concepts. |
| 15 | Perceived Usefulness (PU5) | The supplemental materials helped me learn better. |
| 16 | Perceived Usefulness (PU6) | This AI-enhanced approach is more useful than conventional course materials (e.g., just worksheets). |
| 17 | Ease of Use (PEU1) | It was not difficult for me to learn to access/operate the supplemental materials. |
| 18 | Ease of Use (PEU2) | It only took me a short time to fully know how to use the supplemental materials. |
| 19 | Ease of Use (PEU3) | The content presented in the supplemental materials was easy to understand and follow. |
| 20 | Ease of Use (PEU4) | I quickly learned to use the supplemental materials. |
| 21 | Ease of Use (PEU5) | It was not difficult for me to use the materials during my study time. |
| 22 | Cognitive Load (CL1) | The learning content in the supplemental materials was difficult for me. |
| 23 | Cognitive Load (CL2) | I had to put a lot of effort into understanding the information in the supplemental materials. |
| 24 | Cognitive Load (CL3) | It was troublesome for me to process the information in the supplemental materials. |
| 25 | Cognitive Load (CL4) | I felt frustrated trying to understand the supplemental materials. |
| 26 | Cognitive Load (CL5) | I did not have enough time to review the supplemental materials effectively. |
| 27 | Cognitive Load (CL6) | The way that AI presents content in the supplemental materials caused me a lot of mental effort. |
| 28 | Cognitive Load (CL7) | I needed to put lots of effort into connecting the supplemental materials to my course objectives. |
| 29 | Cognitive Load (CL8) | The logic and flow of the supplemental materials was difficult to follow and understand. |
| 30 | Germane Load | The supplemental materials encouraged me to engage in deeper thinking or reflection. |
| 31 | Germane Load | The supplemental materials help me connect new information to my existing knowledge. |
| 32 | Intrinsic Load | Were there any EAP2 topics that you found particularly challenging? If so, which ones and why? |
| 33 | Coherence Principle | Were there any elements in the supplemental materials that you felt were unnecessary or distracting? If so, please specify. |
| 34 | Qualitative | Do you have any suggestions for improving the supplemental materials for next term's EAP2 students? |
| 35 | | Would you be open to (the instructor) briefly interviewing you to understand your thoughts on supplemental materials? |

**C. Interview Questions**

Please think about the supplemental material that you most frequently use and that you chose above.

1. (TAM: Perceive usefulness; ease of use) Can you explain how the combination of audio, text or visuals in your supplemental material aided your understanding?
2. (TAM and Cognitive Load: Coherence Principle) Can you provide an example of how the length of your supplemental material affected your ability to focus and learn?
3. (Cognitive load) Were there features in the supplemental material that helped you identify important information?
4. (Germane load) How can (the first author) improve your supplemental material to engage students in deeper thinking or reflection? Please explain.
5. (Germane load) How can (the first author) improve your supplemental material to connect new information in (the EAP course) to students' existing knowledge?